\documentclass[a4paper]{toledoStyle}
\usepackage{epsfig}

\begin{document}

\title{Radiative transfer in molecular lines}

\author{A.\,Asensio Ramos\inst{1} \and J.\,Trujillo Bueno\inst{1,3} \and 
  J.\,Cernicharo\inst{2,3} } 

\institute{
  Instituto de Astrof\'\i sica de Canarias, E-38200 La Laguna, Tenerife, Spain
\and 
  Instituto de Estructura de la Materia, Serrano 123, E-28006 Madrid, 
Spain 
\and
  Consejo Superior de Investigaciones Cient\'\i ficas, Spain}

\maketitle 


\begin{abstract}

The highly convergent iterative methods developed by Trujillo Bueno and Fabiani 
Bendicho (1995)
for radiative transfer (RT) applications are generalized
to spherical symmetry with velocity fields. These RT methods are based on
Jacobi, Gauss-Seidel (GS), and SOR iteration and they form the basis of a new 
NLTE multilevel
transfer code for atomic and molecular lines. 
The benchmark tests carried out so 
far
are presented and discussed. The main aim is to develop a number of powerful
RT tools for the theoretical interpretation
of molecular spectra.

\end{abstract}


\section{Introduction}
\label{introd}

Atomic line emission has been extensively used for tracing the physical 
conditions in many
astrophysical plasmas. The relatively high energy difference between the atomic 
energy levels makes this 
diagnostic tool a suitable one for tracing 
the physical conditions in warm and hot media. But cool plamas are not well 
traced 
by atomic lines because the thermal energy is not high enough to populate the 
upper levels of the transitions. Fortunately, FIRST
will allow us to study molecular line emission in greater detail. 
Molecules have 
 very
rich spectra, arising from transitions between the electronic, vibrational and 
rotational levels. Their 
spectra cover the spectral range from radio to optical wavelengths, depending on the type 
of transition.
One has to include very complicated 
molecular systems to be able to model the observations and an extensive forward
RT modeling effort is frequently needed.
With this motivation in mind, we are developing  
a radiative transfer code based on the fast iterative methods
developed by \cite*{tru95} for Cartesian coordinates. Our first step 
was to
generalize these methods to spherical symmetry 
with velocity fields. This new RT tool will help us in the interpretation of 
different kinds of
observations, including ro--vibrational bands in circumstellar envelopes of 
C-rich or O-rich evolved
stars, rotational lines in molecular clouds, molecular emission from the Sun, 
maser emission and many 
others. Another interesting application would be the modeling of maser 
polarization.
  

\section{The state of the art and our approach}
\label{state_art}

The radiative transfer problem requires the self-consistent solution of the rate equations for the 
populations of the molecular levels and the radiative transfer equation. This set of equations
describes a nonlinear and nonlocal problem. Iterative methods are therefore needed. 
Since \cite*{ber79}, most of the 
radiative transfer tools used in molecular radiative transfer have been based on Monte Carlo techniques 
for the solution of the radiative transfer equation and on the $\Lambda$-iteration for the iterative 
solution of the non-LTE problem. The Monte Carlo technique is very powerful for its ability to cope with
complicated geometries, but it has a major 
drawback: its instrinsic random noise. There have been many efforts to reduce the noise, but it
is always present (see, for example,
 \cite{ber79}). On the other hand, the $\Lambda$-iteration scheme
is very easy to implement, but is only useful for optically thin problems. Recently, Accelerated 
Lambda Iteration (ALI) methods have been applied to molecular radiative 
transfer. ALI is a method that requires not much more computational time per iteration than the
$\Lambda$-iteration, but it has a better convergence behavior for optically thick problems. It has 
been extensively used in stellar and solar astrophysics and has recently been
combined with Monte Carlo techniques for molecular RT (\cite{hog00}).

\begin{figure}[ht]
  \begin{center}
  	\begin{tabular}{c@{\quad}c}
    \epsfig{file=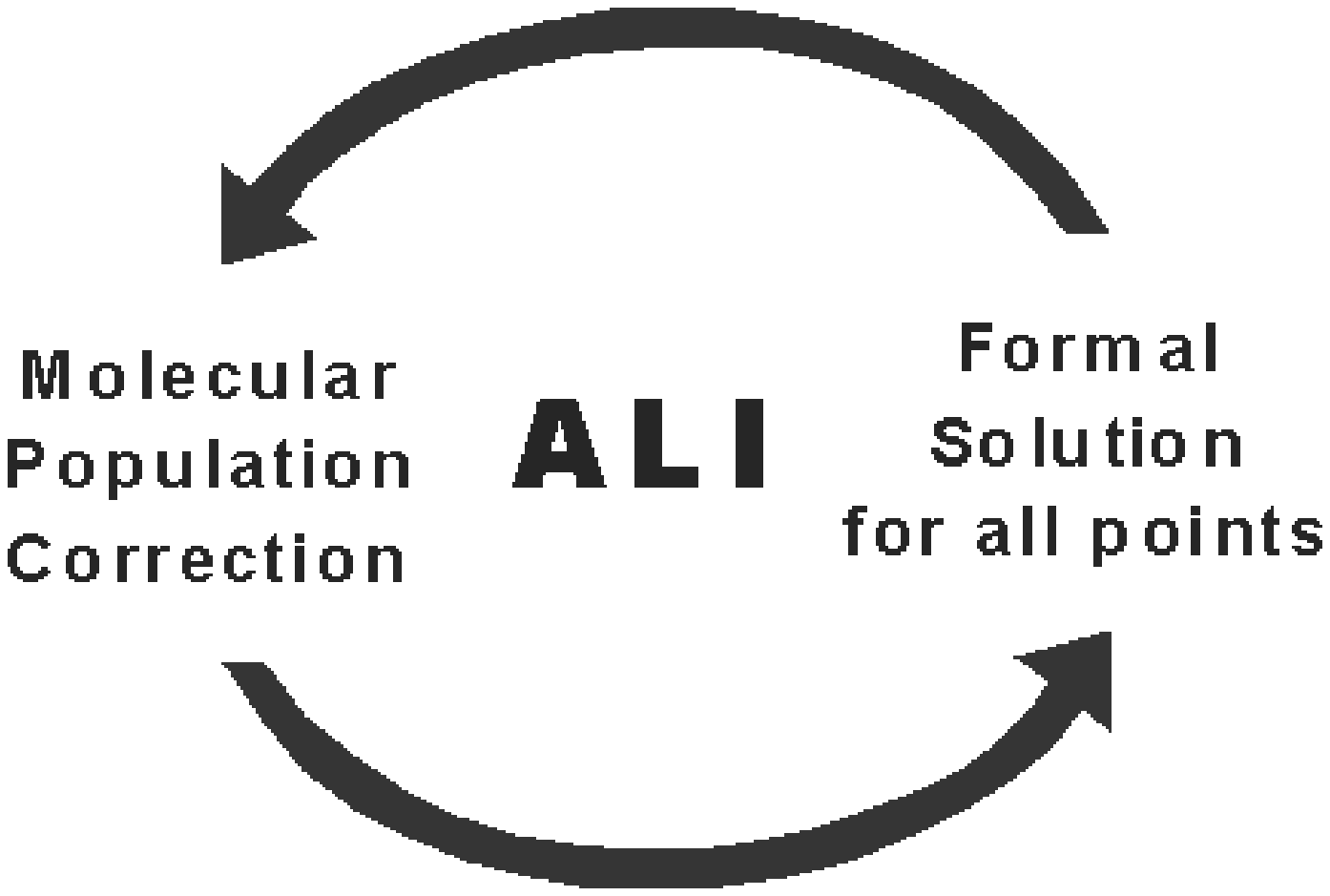, width=4.2cm}
	 \epsfig{file=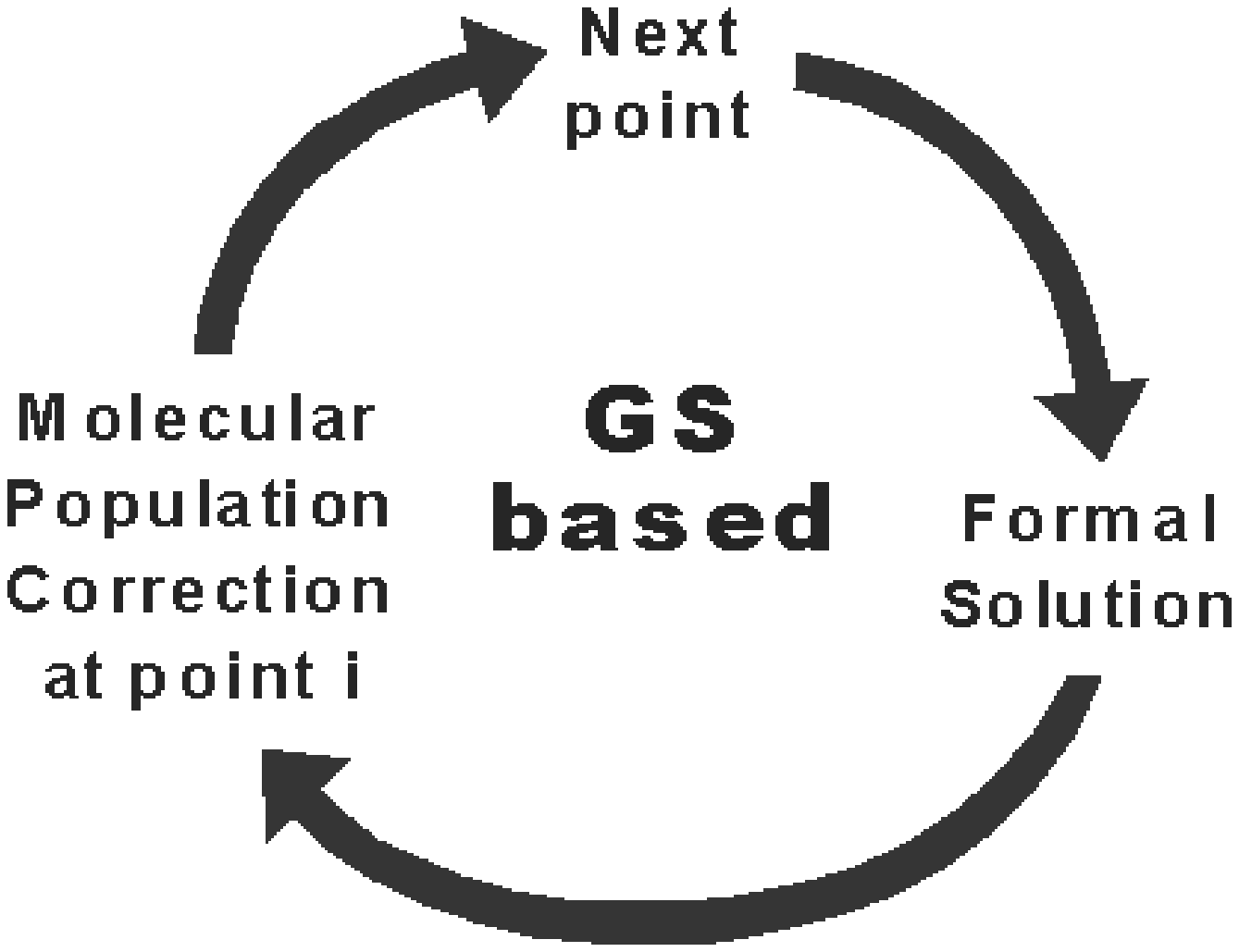, width=4.2cm}
	\end{tabular}
  \end{center}
\caption{ Scheme showing the differences between the ALI iterative method and the GS-based 
method. The figure is explained in the text. }  
\label{fig1}
\end{figure}

Our approach is based on the iterative methods developed by \cite*{tru95}, which
themselves are
 based on the Gauss--Seidel 
scheme and applied to Cartesian coordinates in 1D, 2D and 3D, 
either with or without 
polarization. They allow the solution of the radiative
transfer problem with the same computational time per iteration as ALI, but with an order-of-magnitude improvement
in the number of iterations. The short-characteristics technique (\cite{kun88})
has been chosen as the formal
solver. This has facilitated the implementation of the scheme in a very efficient way (see \cite{tru95}). 
We have generalized
these methods to spherical symmetry with velocity fields. The problem is still one dimensional because
the physical variables  have only radial dependence, 
but angular information has to be achieved
more precisely than for a
 plane--parallel atmosphere in order
to take account of curvature effects. This angular 
information is obtained by means
of solving the RT equation through the impact parameters, as is
usually done. The velocity fields are treated in the 
observer's frame. In Fig. (\ref{fig1}) we show a schematic representation of the difference between
the ALI-based 
and the GS-based iterations. With ALI, when the radiation field is obtained at all the points of the
atmosphere, the statistical equilibrium equations for the molecular population can be solved. On the 
other hand, the main idea behind the GS-based methods is the
fact that when the radiation field is known at one point in the atmosphere, one can write the statistical
equilibrium equations for this point and do the level population correction. When solving the RT equation
to get the radiation field at the next point, we have to take into
account that the population in the
previous point has been improved. This scheme, coded in an efficient way with the aid of the 
short-characteristics formal solver,
can lead to a high improvement in the total number of iterations, while the time per iteration is
virtually the same as with ALI.

\section{Illustrative examples}
\label{examples}
In order to verify that our code is giving reliable results, we have chosen several benchmark
problems whose results have already been published.


\subsection{Bernes's CO cloud}
\label{bernes}

\begin{figure}[ht]
  \begin{center}
  	\begin{tabular}{c@{\quad}c}
    \epsfig{file=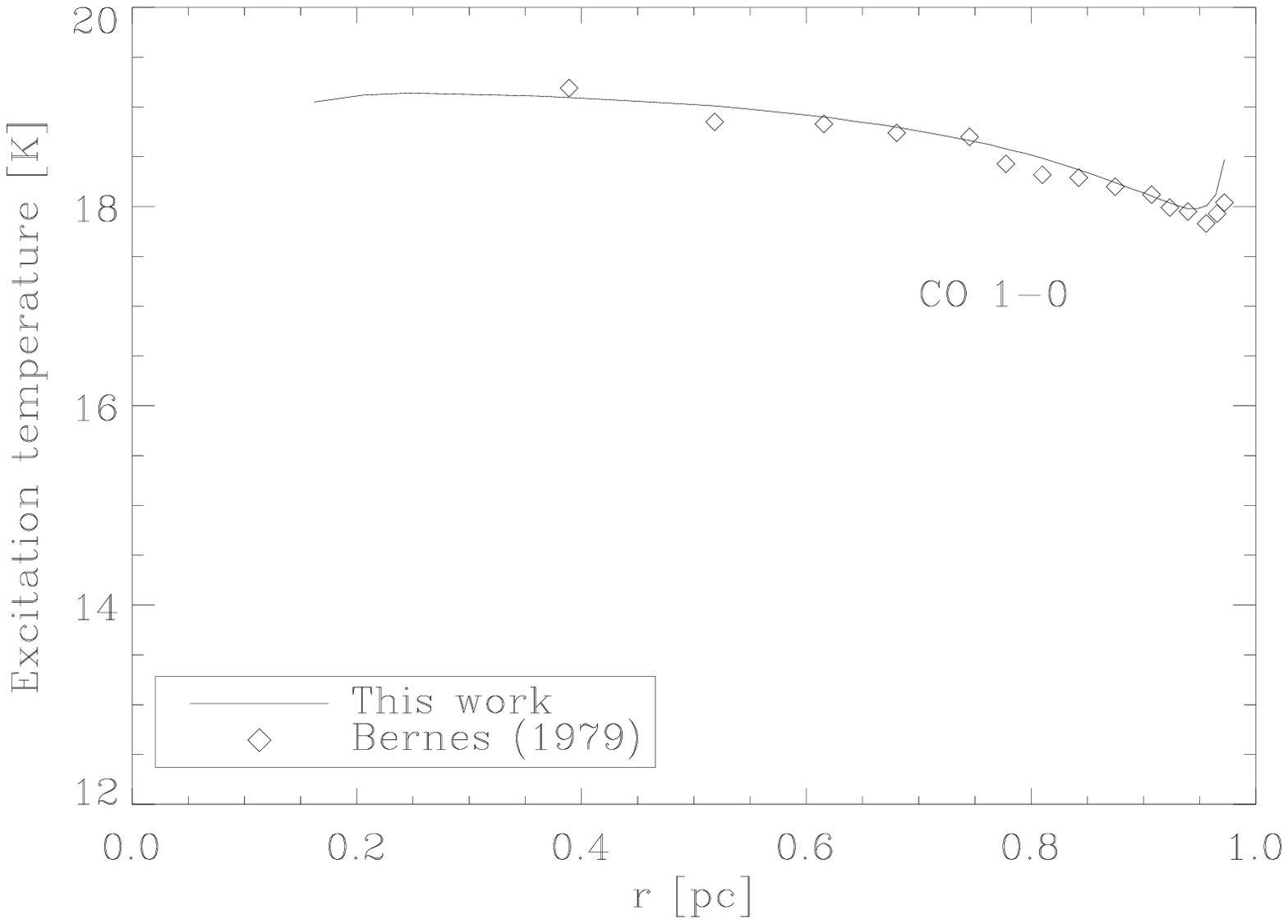, width=4.2cm}
	 \epsfig{file=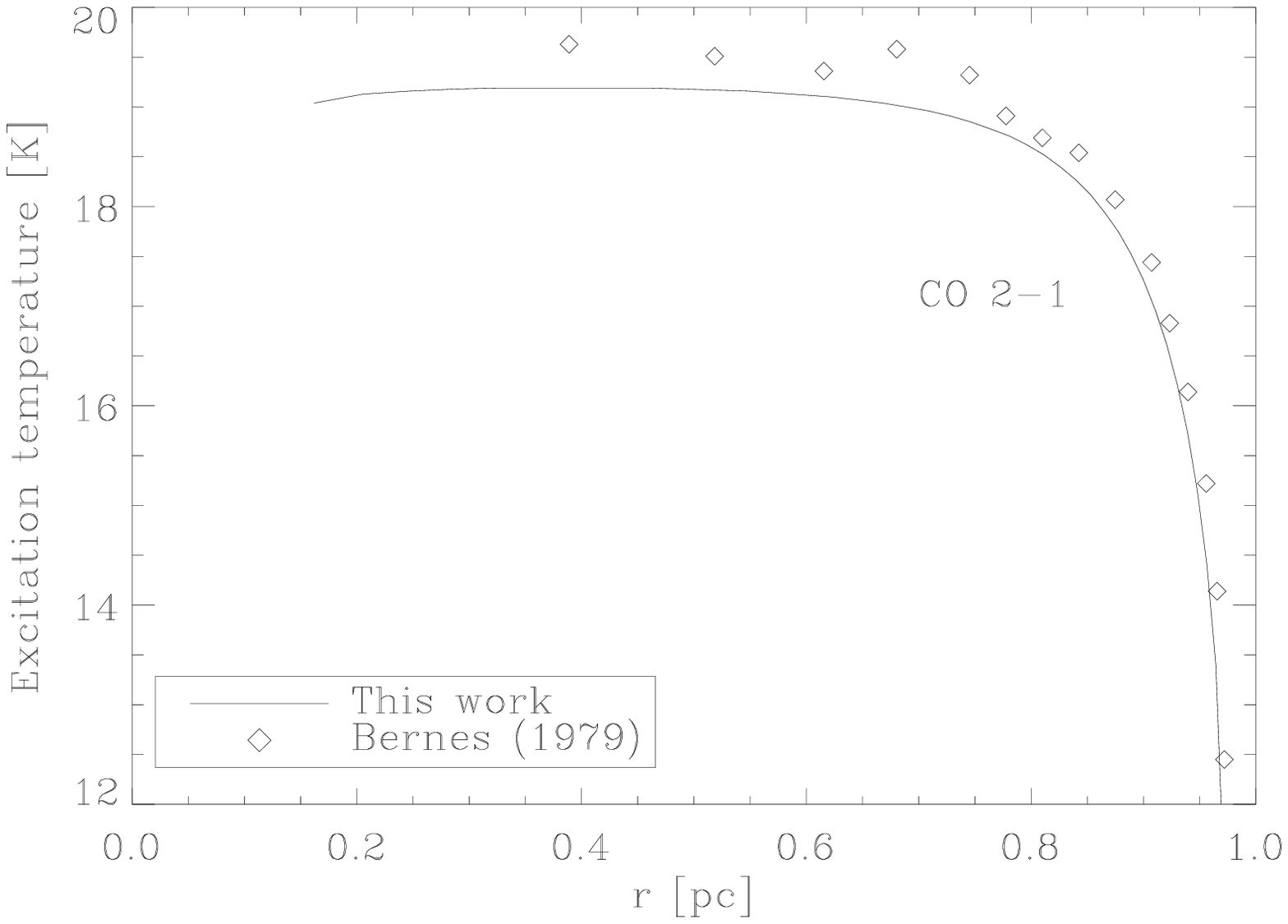, width=4.2cm}
	\end{tabular}
  \end{center}
\caption{ Excitation temperature for the $J=1\rightarrow 0$ and 
$J=2\rightarrow 1$ rotational transitions of CO in the Bernes'
cloud. Comparison between the our results and those of Bernes (1979) are plotted. }  
\label{fig2}
\end{figure}

Let us begin with the CO cloud model used by \cite*{ber79} to introduce his Monte Carlo code. The problem consists
in a constant-density ($n_{{\rm H}_{2}}=2 \times 10^{3}$  cm$^{-3}$), 
constant-temperature ($T=20$~K), 1 pc radius 
infalling cloud with a maximum velocity of 1  km  s$^{-1}$ at the external parts and sampled at 40
radial shells. The CO abundance is $5\times10^{-5}$ and the cloud is illuminated by the cosmic microwave background 
radiation (CMBR) at a temperature of 2.7 K. The CO molecule with the first six rotational levels is used, 
taking into account that the same  collisional rates used by Bernes in his calculation have to be used to
get 
similar results. In Fig. (\ref{fig2}) we show the excitation temperature for the transitions $J=1\rightarrow 0$ and $J=2\rightarrow 1$ 
through the cloud. We show the results obtained with our code and that obtained by Bernes using his Monte Carlo technique. 
The intrinsic noise of the Monte Carlo scheme can clearly be appreciated in the figures.

\subsection{Leiden benchmark test}
\label{leiden}
\begin{figure}[ht]
  \begin{center}

	 \begin{tabular}{c@{\quad}c}
      \epsfig{file=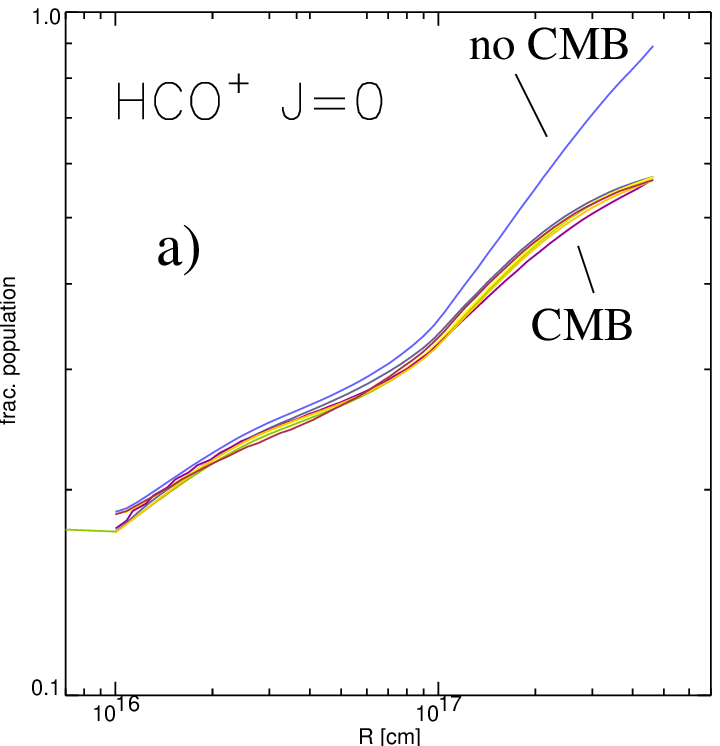, width=4.2cm, height=3.2cm}
	   \epsfig{file=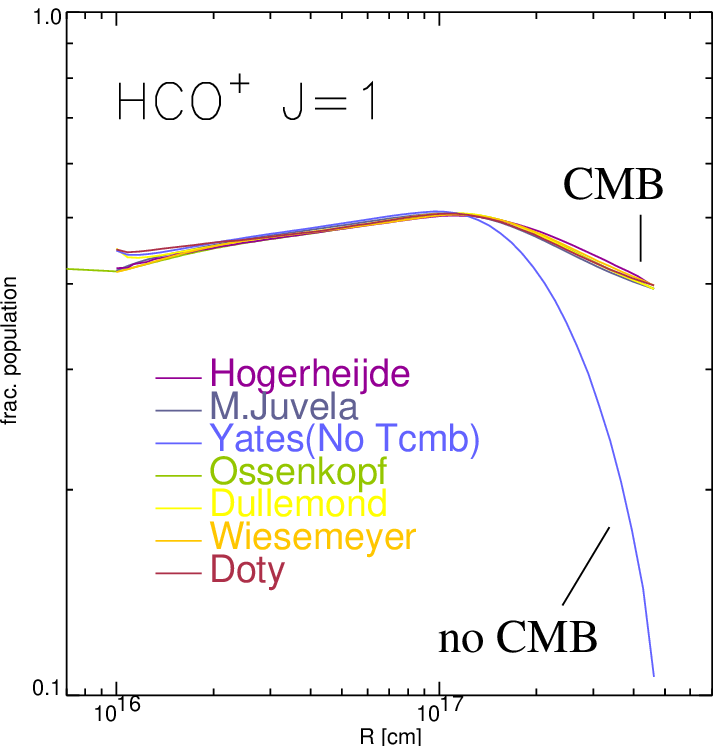, width=4.2cm, height=3.2cm}
	 \end{tabular}
	 
	 \begin{tabular}{c@{\quad}c}
      \epsfig{file=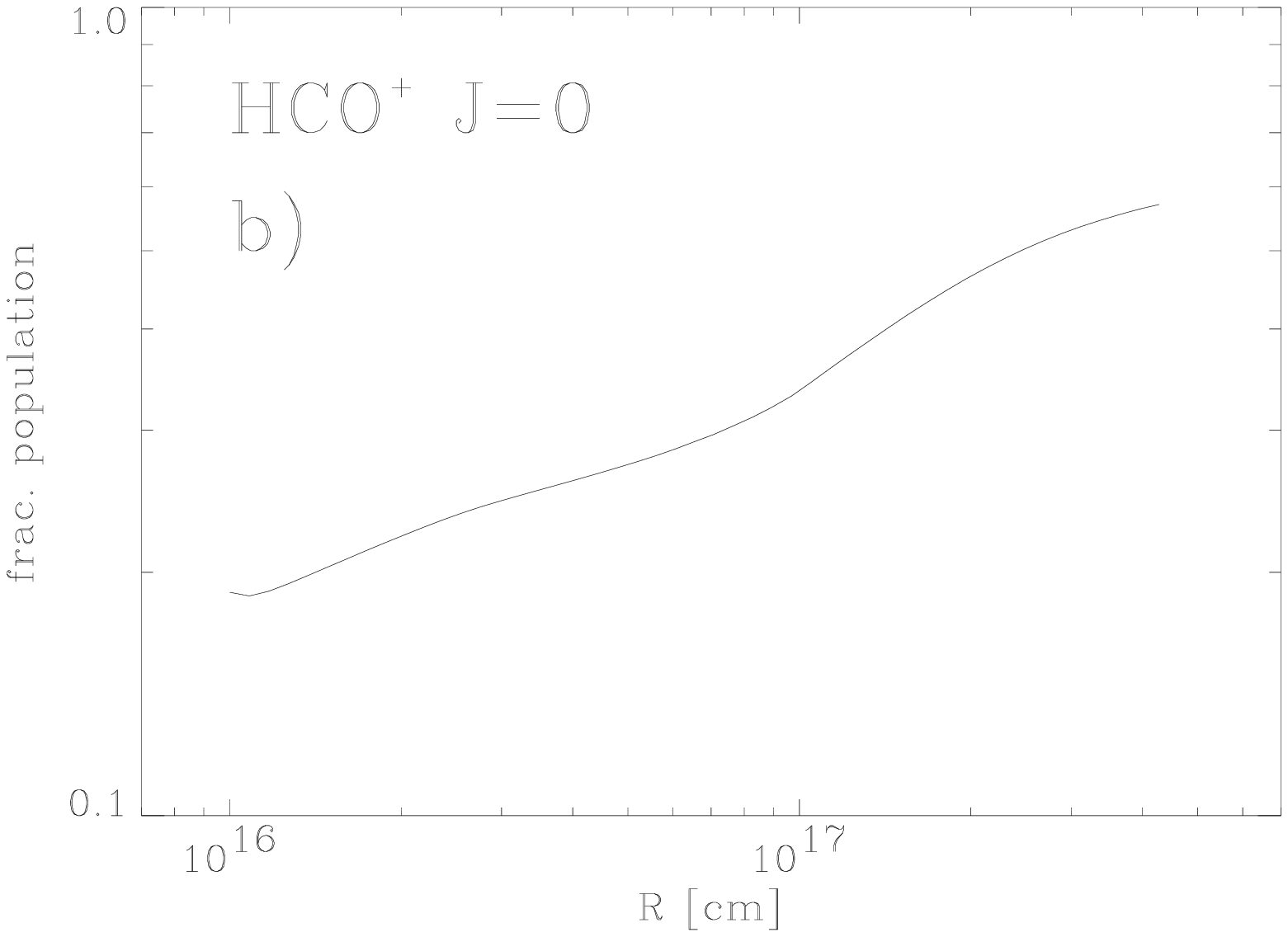, width=4.2cm, height=3.2cm}
	   \epsfig{file=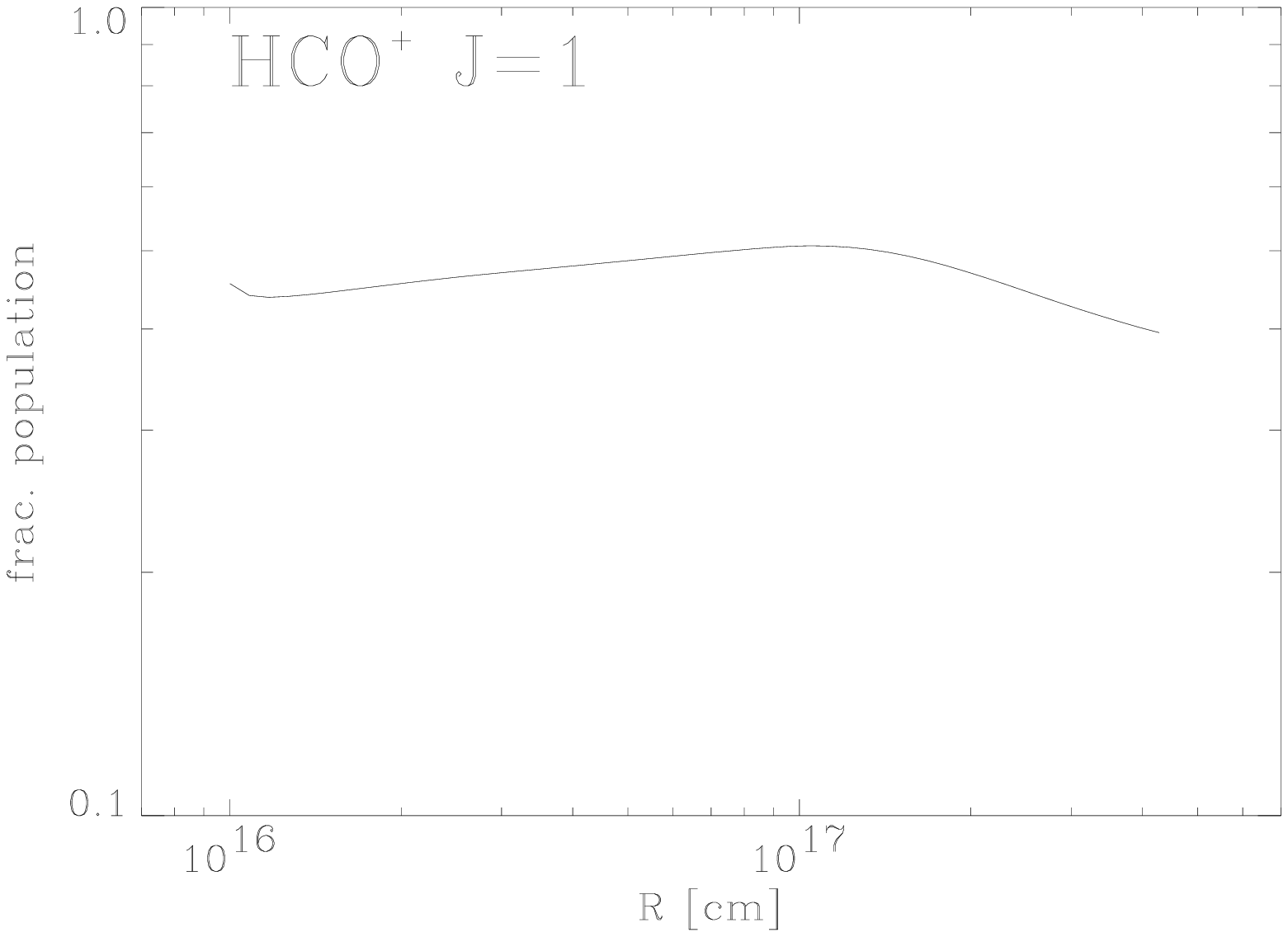, width=4.2cm, height=3.2cm}
	 \end{tabular}
	 
	 \begin{tabular}{c@{\quad}c}
      \epsfig{file=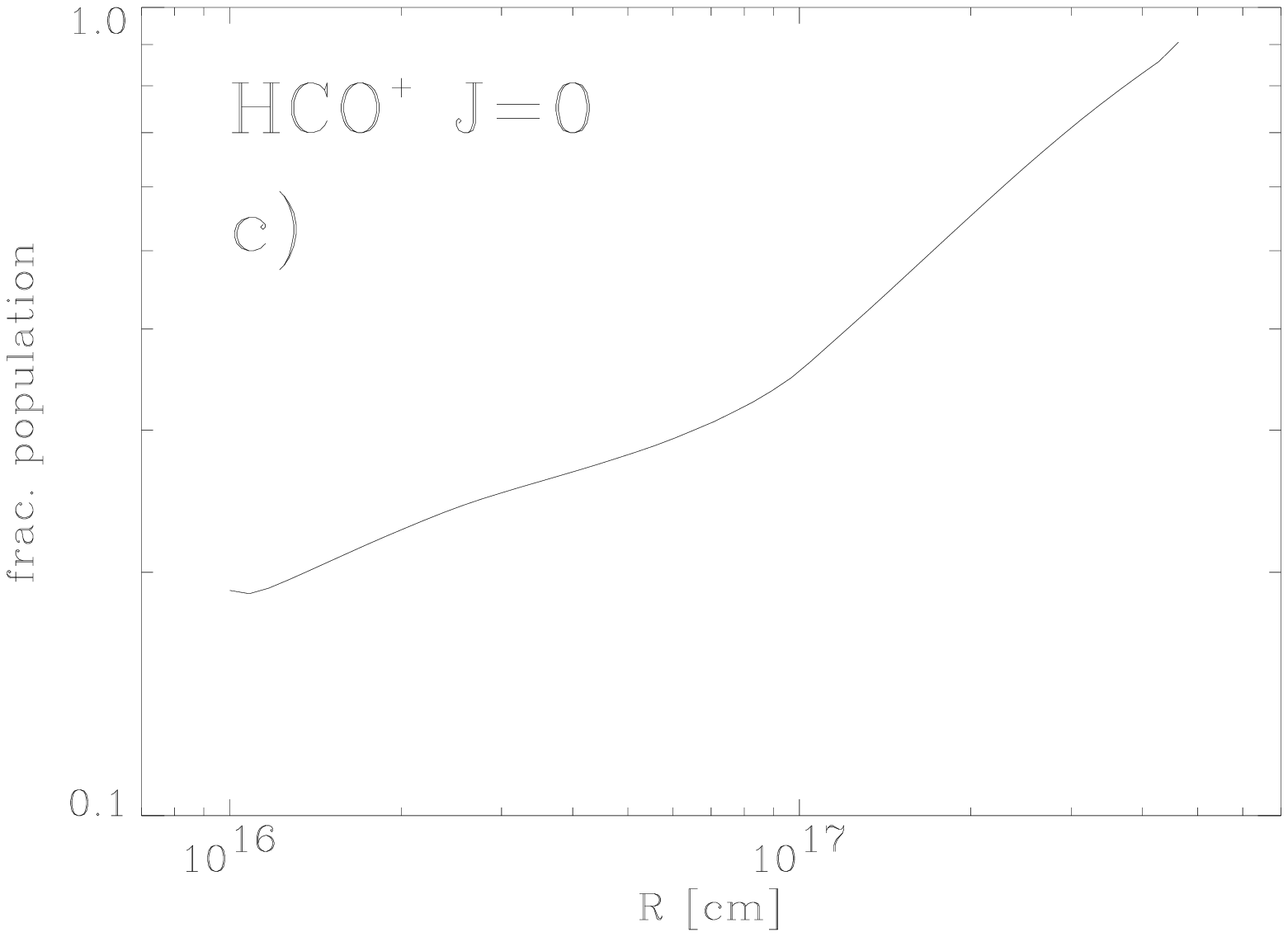, width=4.2cm, height=3.2cm}
	   \epsfig{file=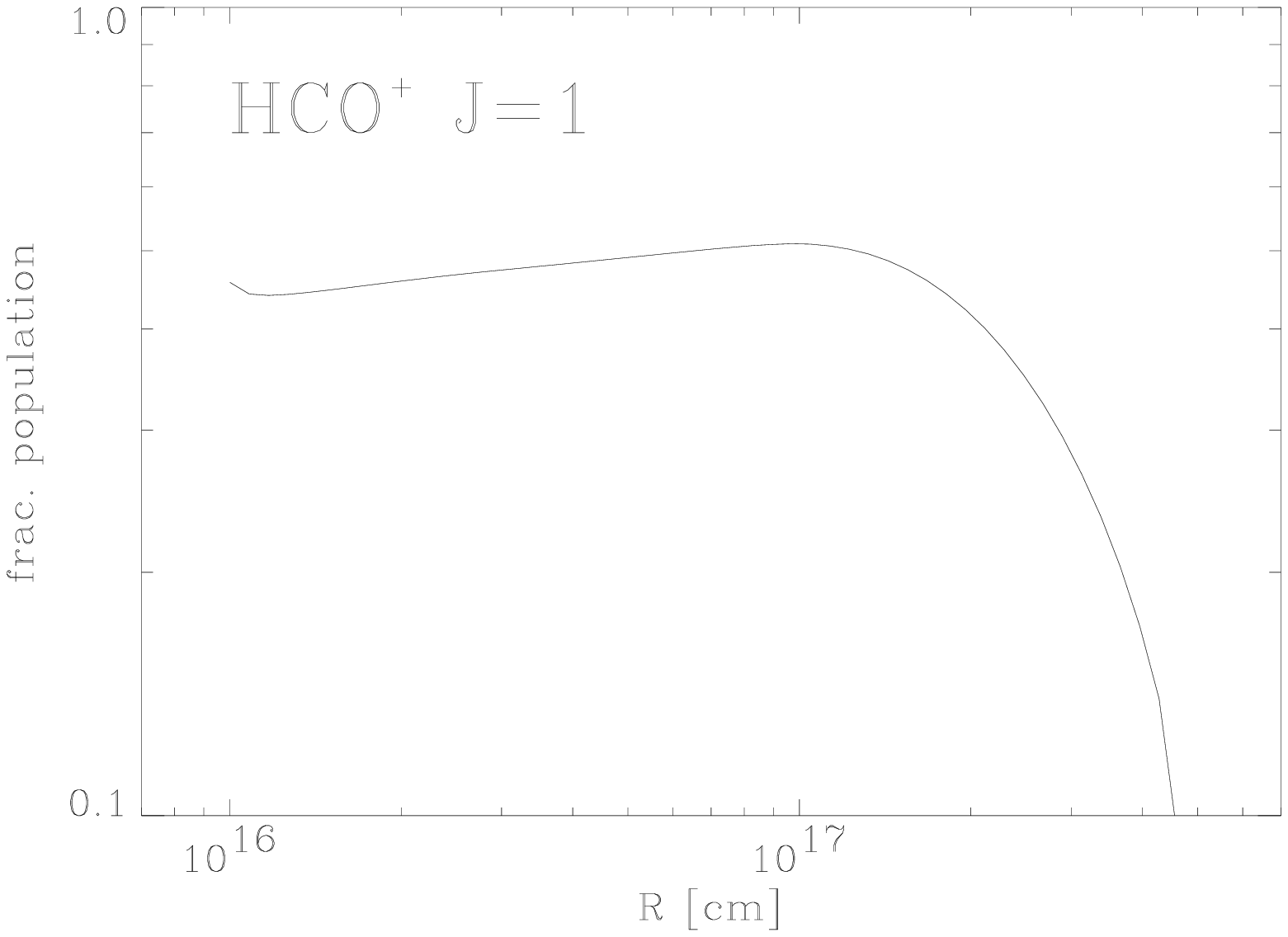, width=4.2cm, height=3.2cm}
	 \end{tabular}

  \end{center}
\caption{ Fractional population of the first two rotational levels of HCO$^{+}$. Panel a) represents the
results obtained with seven different codes, while panels b) and c) represent the results obtained with
our code by including (b) and excluding (c) the CMBR. }  
\label{fig3}
\end{figure}

A number of useful test cases became  available after the 1999 workshop on 
{\it Radiative Transfer in Molecular
Lines} at the Lorentz Center of Leiden University\footnote{http://www.strw.leidenuniv.nl/$\sim$radtrans}. 
These are intended for the testing of newly developed molecular RT codes
 against already existing ones. 
Although every
test problem has been solved with our multilevel NLTE code and  good agreement
obtained,
we   show only some of the results.
The model describes a collapsing cloud similar to that
 described by \cite*{shu77}, where the first
21 rotational levels of HCO$^{+}$ (from $J=0$ to $J=20$) are taken into account in the non-LTE calculation. The
molecular abundance is $\left [ \rm{HCO}^{+} \right ]=10^{-9}$, so lines are only slightly optically thick
($\tau < 10$). The cloud is sampled logarithmically at 50 depth points and is externally illuminated by 
the CMBR  at 2.728 K. Results for $J=0$ and $J=1$ are
given if Fig. (\ref{fig3}a) for the different codes used in the test and in Fig. (\ref{fig3}b)
corresponding to our code. In these plots 
we represent the fractional population of each level, which can be written as 
$f=n_{\rm level}/n_{\rm total}$.

\begin{figure}[!ht]
  \begin{center}
    
	 \begin{tabular}{c@{\quad}c}
      \epsfig{file=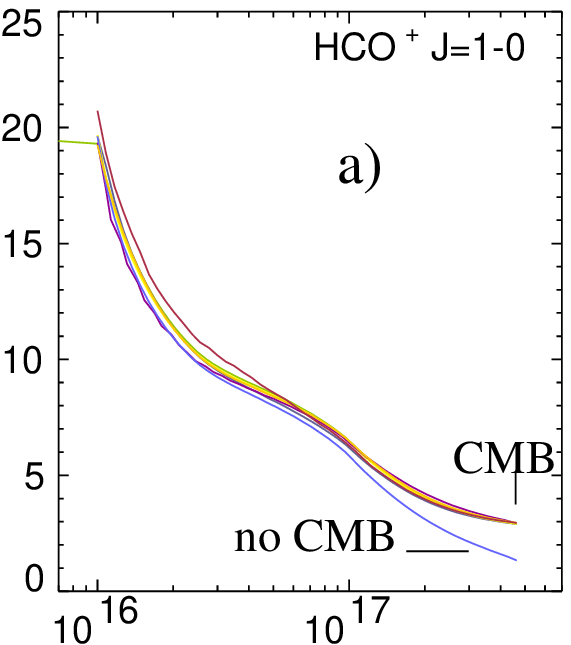, width=4.2cm, height=3.2cm}
	   \epsfig{file=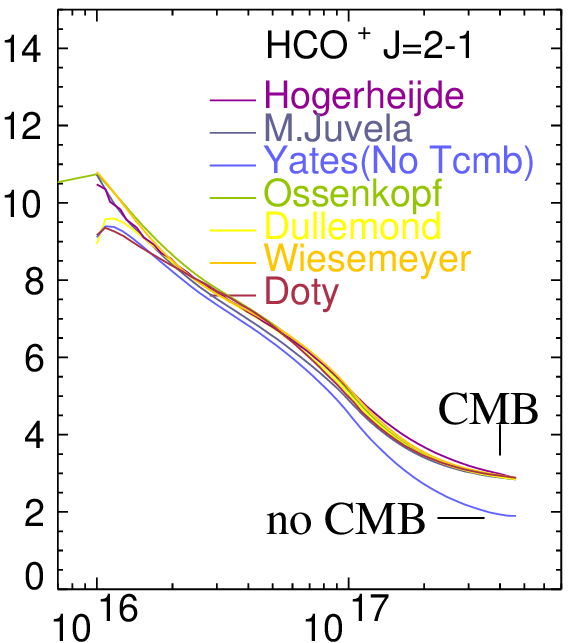, width=4.45cm, height=3.2cm}
	 \end{tabular}

	 \begin{tabular}{c@{\quad}c}
      \epsfig{file=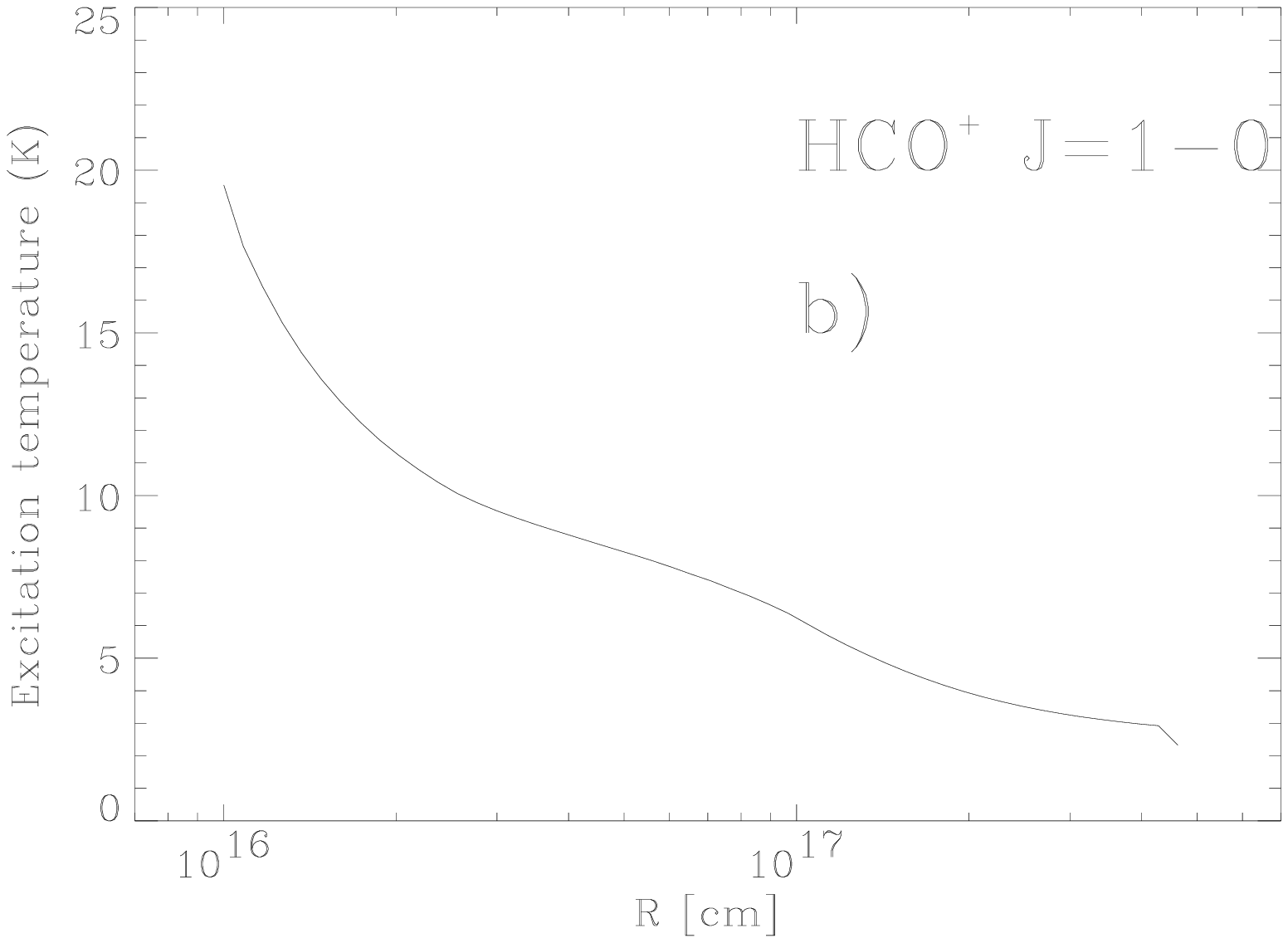, width=4.2cm, height=3.2cm}
	   \epsfig{file=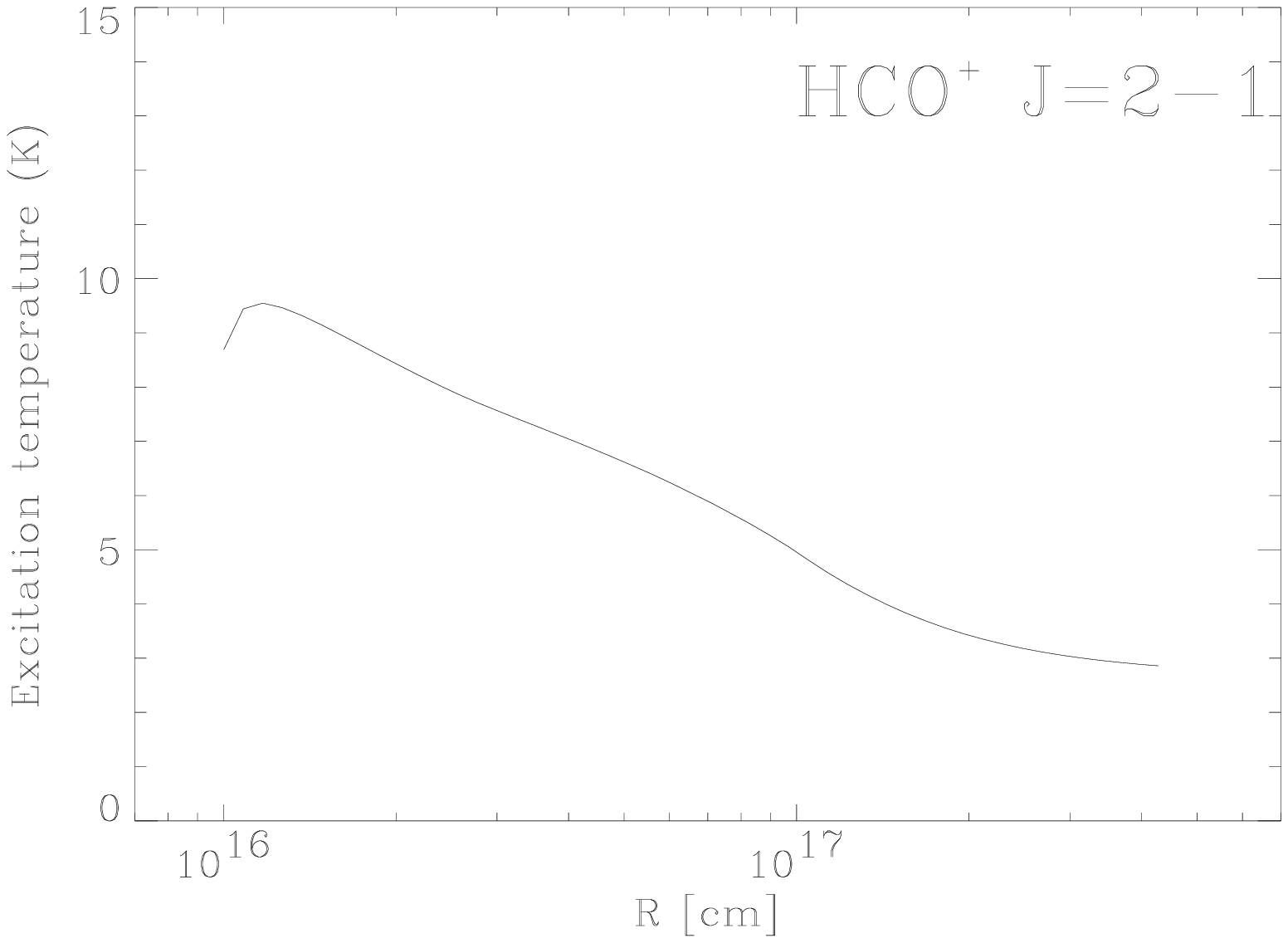, width=4.2cm, height=3.2cm}
	 \end{tabular}
	 
	 \begin{tabular}{c@{\quad}c}
      \epsfig{file=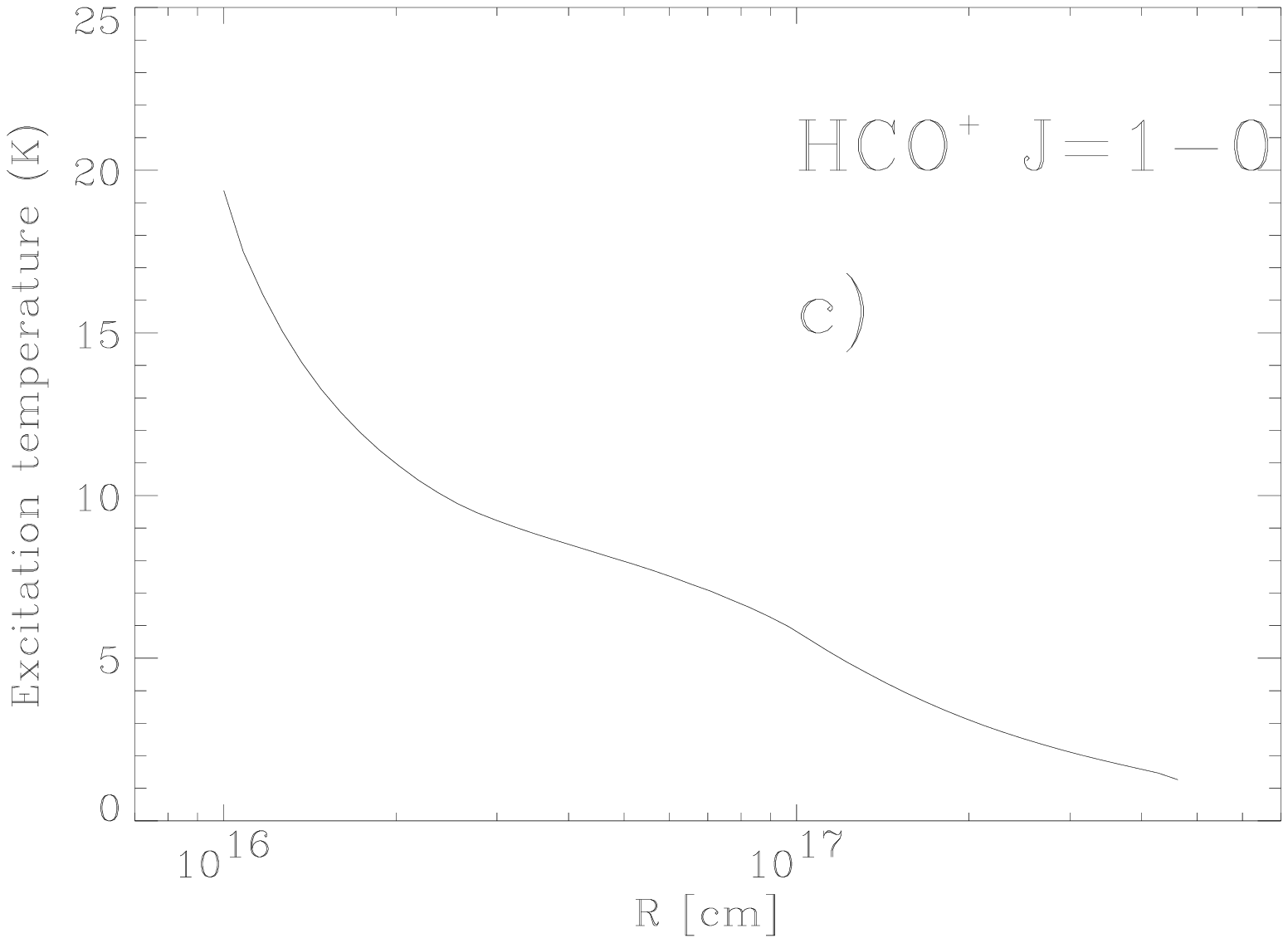, width=4.2cm, height=3.2cm}
	   \epsfig{file=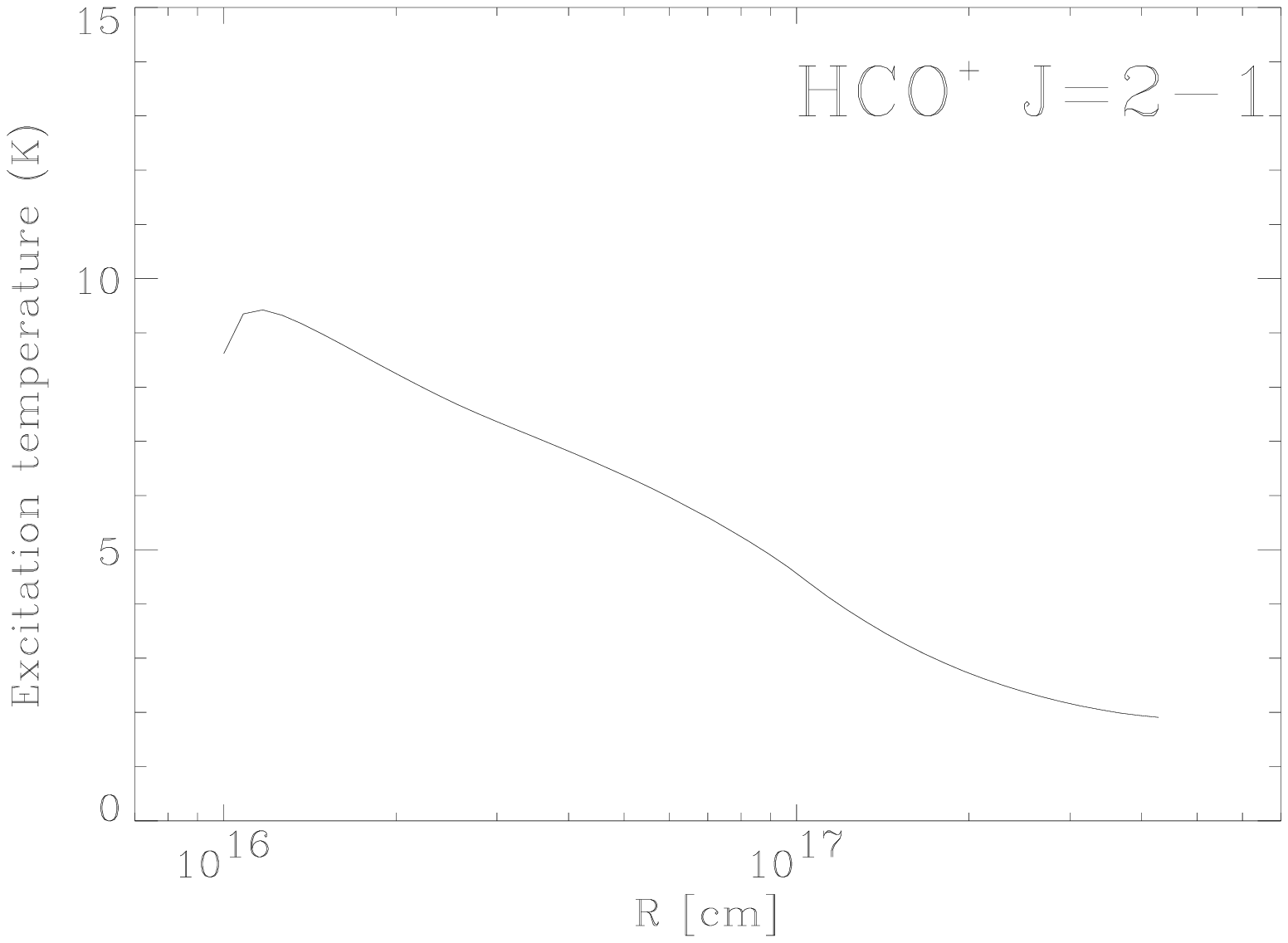, width=4.2cm, height=3.2cm}
	 \end{tabular}	 

  \end{center}
\caption{ Excitation temperature for the rotational transitions between the three lowest levels of 
HCO$^{+}$. Panel a) represents the
results obtained with seven different codes, while panels (b) and (c) represent the results obtained with
our code, either including the CMBR (b) or not (c). }  
\label{fig4}
\end{figure}

We see that the results agree. Although it cannot be seen in our plots, most 
of the HCO$^{+}$ in the inner parts of the cloud 
is in the lowest four rotational levels, because the kinetic temperature is relatively high ($T < 20$~K) 
and there is energy in the medium to populate the higher levels. On the other hand, in the external zones 
of the cloud, almost 60$\%$ of the HCO$^{+}$ is at the ground level ($J=0$), and the remaining 40$\%$ is in the
$J=1$ level due to the lower kinetic temperature, which is not able to  populate the higher levels efficiently. 
As can be seen in Fig. (\ref{fig3}a), there is one curve which is different from the others. This is
caused by not having included the CMB radiation as the outer boundary condition and assuming that the cloud
is not externally illuminated. This turns out to be an extra test for our code, 
and the results for this particular situation are shown in Fig. (\ref{fig3}c). 
Agreement is also obtained for the remaining levels, and one can see that there is no
significant difference between the results in the inner parts of the cloud. However, a totally different result 
is obtained in the external zones, where the ground level is the only one populated with $\sim$90\% of the
total abundance. Although the kinetic temperature at the outer envelopes of the cloud
is still able to populate higher levels, non-LTE effects produce this underpopulation of the 
higher levels.
Excitation temperature is also plotted in Fig. (\ref{fig4}), for the two transitions $J=1\rightarrow 0 $ and the $J=2\rightarrow 1$. Also
in Figs. (\ref{fig4}b) and (\ref{fig4}c) the results obtained with our code are also plotted, either including 
the CMBR as a boundary condition or not, respectively. 
The results are also comparable to those obtained 
by different
codes in both cases. There is a little more dispersion in this result than in that for  fractional population,
but this could be due to the fact that the majority of the codes are based on Monte Carlo schemes, which, 
although they have variance reduction techniques, have an intrinsic random noise that could produce
these effects.

\begin{figure}[ht]
  \begin{center}
    \epsfig{file=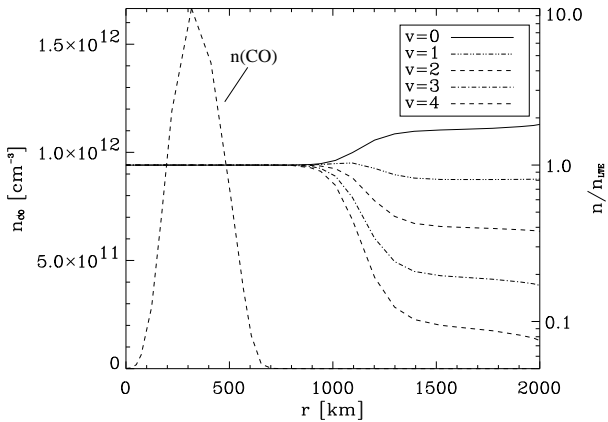, width=8cm}
  \end{center}
\caption{ Departure coefficients (right axis) and CO abundance (left axis) in a non-LTE calculation
for the $\Delta v=1$ band at 4.7 $\mu$m in a quiet-Sun model. The first five vibrational levels ($v=0$ to
$v=4$) with 21 rotational levels ($J=0$ to $J=20$) within each one are included. Note that LTE is obtained for
CO in the line-formation region.}  
\label{fig5}
\end{figure}


\subsection{CO in the Sun}
\label{co_sun}

We have solved the non-LTE problem for the $\Delta v=1$ ro--vibrational band of CO at 4.7 $\mu$m. 
The vibrational constant for CO is $\omega_{0}=2143$~cm$^{-1}$ and the rotational constant for the 
vibrational ground state is $B_{0}=1.923$~cm$^{-1}$. Since $\omega_{0} \gg B_{v}$, it follows that
successive rotational levels within one vibrational state are much closer in energy than similar 
rotational levels in successive vibrational states. It can be shown that spontaneous radiative decay rates
for pure rotational transitions are much lower than collisional rates (see,
for example, \cite{tho73}), 
so one may assume without many
problems that the populations of the rotational levels within a vibrational state are given by  
Boltzmann statistics. This assumption greatly simplifies the problem as 
shown by \cite*{uit00} for the same CO
problem, because the number of unknowns is 
reduced from the total number of levels (the
number of vibrational levels $\times$ number of rotational
levels within each vibrational state) to the total number of vibrational 
levels.
However, we have solved the whole problem without making
 this assumption and including the first five vibrational 
levels 
(from $v=0$ to $v=4$) and 21 rotational levels within each vibrational one (from $J=0$ to $J=20$). A quiet-Sun
model atmosphere has been chosen (\cite{val81}) and the molecular abundance has been calculated 
in this model assuming chemical equilibrium. As shown in Fig. (\ref{fig5}), the CO abundance peaks at 
$\sim 300$~km above the bottom of 
the photosphere. This figure also shows the departure
coefficients for all the vibrational levels included. These departure coefficients are calculated as usual, but
taking into account the total population of each vibrational level, which can be obtained by summing over the
rotational levels inside each vibrational one: $b = \sum_J{n_{\rm NLTE}(v,J)} / \sum_J{n_{\rm LTE}(v,J)}$. We see that
LTE is obtained in the line-formation region below $\sim 800$~km, which is the zone where most of the 
CO is formed. This partially
 confirms the results that can be obtained by comparison of the radiative and
collisional transition rates.


\section{Conclusions}
\label{conclus}
We have generalized very efficient iterative methods for the solution of the molecular radiative transfer 
problem to 
spherical geometry with velocity fields. The problem is still one-dimen\-sional, but more angular information
is required in comparison to the plane--parallel case, so the total computation time is larger. Velocity
fields are treated in the observer's frame, so velocity fields have to be limited to several times the thermal 
velocity in the medium if we want to have a tractable frequency quadrature. This limitation is only a 
computational problem and not a true limitation of the method.
Such a fast solution of the non-LTE problem allows the solution of more complicated situations, where 
larger molecular models can be used. It is known that there can be many different pumping processes in 
molecular radiative transfer---very important for the interpretation of 
masers---and a correct model 
including all the possible important levels is crucial for the interpretation of observations. 
On the other hand, the advantage of getting the solution of the non-LTE problem in only a few iterations leads 
to great advantages. This makes it possible to improve the adjusting
of all the physical parameters in the model one is using to interpret the observations, because much more
extensive forward modeling is now possible.
Finally, the fast solution of the radiative transfer problem allows us
 to introduce the transfer of polarized
radiation with the aid of the density matrix theory (see the review by \cite{tru01}). This could lead to the 
self-consistent solution of the maser polarization problem, which could make it possible to explain radio 
observations of masers such as SiO.

\end{document}